# The CAPIRE Curriculum Graph: Structural Feature Engineering for Curriculum-Constrained Student Modelling in Higher Education

Hugo Roger Paz
PhD Professor and Researcher Faculty of Exact Sciences and Technology National University of Tucumán
Email: hpaz@herrera.unt.edu.ar
ORCID: https://orcid.org/0000-0003-1237-7983

## Abstract

Curricula in long-cycle programmes are usually recorded in institutional databases as linear lists of courses, yet in practice they operate as directed graphs of prerequisite relationships that constrain student progression through complex dependencies. This paper introduces the **CAPIRE Curriculum Graph**, a structural feature engineering layer embedded within the CAPIRE framework for student attrition prediction in Civil Engineering at Universidad Nacional de Tucumán, Argentina. We formalise the curriculum as a directed acyclic graph, compute course-level centrality metrics to identify bottleneck and backbone courses, and derive nine structural features at the student–semester level that capture how students navigate the prerequisite network over time. These features include backbone completion rate, bottleneck approval ratio, blocked credits due to incomplete prerequisites, and graph distance to graduation. We compare three model configurations—baseline CAPIRE, CAPIRE plus macro-context variables, and CAPIRE plus macro plus structural features—using Random Forest classifiers on 1,343 students across seven cohorts (2015–2021). While macro-context socioeconomic indicators fail to improve upon the baseline, structural curriculum features yield consistent gains in performance, with the best configuration achieving overall Accuracy of 86.66% and F1-score of 88.08% and improving Balanced Accuracy by 0.87 percentage points over a strong baseline. Ablation analysis further shows that all structural features contribute in a synergistic fashion rather than through a single dominant metric. By making curriculum structure an explicit object in the feature layer, this work extends CAPIRE from a multilevel leakage-aware framework to a curriculum-constrained prediction system that bridges network science, educational data mining, and institutional research.

**Keywords:** CAPIRE framework; curriculum analytics; curriculum graph; structural feature engineering; centrality; educational data mining; interpretable machine learning.

# 1. INTRODUCTION

## 1.1. Problem Setting and Emerging Context

University curricula, particularly in engineering and other STEM programmes, are not ordered lists of independent courses but structured networks of prerequisite relationships that constrain student progression through complex dependencies. Each course functions as a node within a directed graph in which edges represent mandatory prerequisite constraints, creating a topology that fundamentally shapes the feasible space of student trajectories. Institutional analytics, however, often collapse this topology into simple indicators—total credits, cumulative grade point averages, or binary completion flags—thereby discarding the relational structure that governs how students actually navigate towards graduation.

Recent advances in curriculum analytics have begun to address this gap. As a subfield of learning analytics, curriculum analytics leverages institutional data to provide evidence-based insights for curriculum improvement, student success optimisation, and programme-level quality assurance (De Silva et al., 2024; Hilliger et al., 2023). A growing body of work shows that representing curricula as directed graphs—variously termed course–prerequisite networks or curriculum networks— enables rigorous structural analysis that can reveal critical bottlenecks, identify foundational backbone courses, and quantify curriculum complexity through graph-theoretic metrics (Aldrich, 2015; Stavrinides & Zuev, 2023; Yang et al., 2024, 2025). These structural properties have been linked to student outcomes, including time-to-degree, dropout risk, and academic performance trajectories.

In parallel, educational data mining and learning analytics have developed increasingly sophisticated predictive models for student success, drawing on clickstream logs, learning management system interactions, assessment histories, and demographic attributes (Baker & Inventado, 2014; Romero & Ventura, 2020). More recent work explores graph neural networks for student modelling, recognising that educational ecosystems—students, courses, prerequisite structures, and learning resources—naturally form heterogeneous networks in which relational information can enhance predictive power (Li et al., 2022; Albreiki et al., 2023; Liu et al., 2025). Nonetheless, most graph-based approaches in education focus on knowledge tracing or content recommendation rather than on integrating explicit curriculum structure into student attrition models.

The CAPIRE framework has shown that student attrition in engineering is a multilevel phenomenon, emerging from interactions between individual trajectories at the micro level, cohort dynamics at the meso level, and institutional context at the macro level (Perez-Rasetti et al., 2023). Previous work within CAPIRE formalised a leakage-aware feature architecture in which temporal ordering and hierarchical structure are preserved so that prediction at semester $t$ uses only information available at or before *t*–1. This temporal discipline prevents information leakage that would artificially inflate model performance while rendering the system unusable for prospective early-warning interventions.

Despite these advances, a fundamental misalignment persists between how curricula are designed and how they are analysed. Curricula are engineered as networks of constraints embodying pedagogical theory about knowledge sequencing, cognitive load management, and skill scaffolding, yet they are typically analysed as flat feature spaces. The prerequisite graph—which captures the structural logic of the curriculum—remains implicit background rather than an explicit object in the modelling layer. This discrepancy limits both the interpretability and the pedagogical actionability of predictive models, because structural properties that may drive attrition risk are not represented as features available to the model.

### 1.2. Motivation and Research Gap

The present work is motivated by three complementary observations from recent literature and from CAPIRE's empirical findings. First, curriculum network analysis has demonstrated that graph-theoretic centrality measures, particularly betweenness centrality, can identify bottleneck courses that act as critical control points in student flow through the curriculum (Aldrich, 2015; Yang et al., 2024). Courses with high betweenness lie on many shortest paths between other courses, so failure to complete them blocks access to large portions of the curriculum. Yet this structural criticality is rarely translated into student-level features for predictive modelling.

Second, while graph neural networks enable end-to-end learning on graph-structured data, they pose interpretability challenges in educational settings where stakeholders—faculty, advisors, and administrators—require transparent explanations for predictions to support actionable interventions (Guleria & Sood, 2023; Swamy et al., 2023). Recent work in explainable artificial intelligence for education emphasises the need for white-box or inherently interpretable models in which feature importance can be traced to pedagogically meaningful constructs (Pereira et al., 2021). Structural features derived from curriculum graphs offer a middle ground between interpretability and predictive power: they summarise network properties while remaining explicit numerical attributes that can be inspected, discussed, and acted upon.

Third, temporal dynamics in student trajectories—including non-linear engagement patterns, time-varying risk profiles, and longitudinal progression sequences—have been shown to strongly influence outcomes (López-Pernas & Saqr, 2024; Cannistrà et al., 2024). However, most curriculum network analyses remain static, treating the graph as fixed rather than as dynamically experienced by students who traverse it over time. Integrating curriculum structure with observed student trajectories at each semester provides a temporally grounded view of how structural constraints interact with individual progression.

Within CAPIRE, a deterministic Program Evaluation and Review Technique analysis of the Civil Engineering curriculum identified a theoretical critical path and highlighted courses with high schedule pressure. This analysis, however, assumed ideal progression without empirical duration distributions, remained disconnected from real student trajectories, and did not generate reusable features for predictive models. Collectively, these limitations point to a research gap: curriculum structure is known to matter, but it is not yet operationalised as a systematic feature layer within multilevel attrition prediction frameworks.

**1.3. Research Objectives and Contributions**

This paper introduces the **CAPIRE Curriculum Graph** as a structural feature engineering layer embedded within the CAPIRE framework for student attrition prediction in Civil Engineering at Universidad Nacional de Tucumán, Argentina. Our work makes four principal contributions to curriculum analytics and educational data mining.

First, we provide explicit curriculum graph construction by formalising the Civil Engineering plan as a directed curriculum graph with fifty-three prerequisite edges connecting thirty-four courses. We compute course-level centrality metrics—including degree, betweenness, closeness, and eigenvector centrality—to characterise the structural role of each course within the network. Bottleneck courses are identified using betweenness centrality, and backbone courses are defined as the union of all courses on any shortest path between entry-level and capstone nodes.

Second, we derive nine structural features at the student–semester level. These features include structural credits approved, backbone completion rate, bottleneck approval ratio, blocked credits, distance to graduation measured as shortest-path length, number of prerequisites met, mean in-degree and out-degree of approved courses, and module diversity. Each feature captures a distinct dimension of how students position themselves within the curriculum network as they progress through their studies.

Third, we integrate structural features with CAPIRE and macro-context variables. The structural features are embedded into CAPIRE's existing multilevel architecture alongside baseline trajectory features and macro-context socioeconomic indicators. We evaluate three configurations—baseline, baseline plus macro variables, and baseline plus macro plus structural features—using Random Forest classifiers with stratified cross-validation, and show that structural features deliver modest but meaningful improvements in Accuracy, Balanced Accuracy, and F1-score.

Fourth, we implement a reproducible pipeline and an ablation-based interpretability analysis. The entire workflow—from curriculum graph construction through structural feature derivation to model training—is implemented in Python and made reusable. An ablation study in which each structural feature is removed in turn shows that all features have comparable importance and operate synergistically rather than through a single dominant metric.

These contributions are organised around three research questions:

- **RQ1:** How does the Civil Engineering curriculum's prerequisite graph structure (backbone, bottlenecks, and path lengths) characterise the constraints faced by students as they progress?

- **RQ2:** To what extent do structural curriculum features improve predictive performance over a strong baseline CAPIRE model with macro-context variables?

- **RQ3:** What do feature importance and ablation analyses reveal about the relative and collective contribution of structural features to attrition prediction?

By answering these questions, the paper extends CAPIRE from a multilevel leakage-aware analytics framework to a curriculum-constrained modelling system in which structural properties of the curriculum become first-class objects in the prediction layer.

## 2. RELATED WORK

The present work draws on five interconnected strands of research: curriculum analytics and graph-based curriculum modelling; graph neural networks and graph-based machine learning in education; feature engineering for student success prediction; temporal and longitudinal modelling of student trajectories; and explainability in educational machine learning. A sixth strand, a previous Program Evaluation and Review Technique analysis conducted within the CAPIRE project, functions as a conceptual predecessor to the present structural feature layer. This section reviews each strand and highlights the specific gaps that motivate the CAPIRE Curriculum Graph.

### 2.1. Curriculum Analytics and Graph-Based Curriculum Modelling

Curriculum analytics has emerged as a specialised subfield within learning analytics that focuses on the design, structure, and governance of academic programmes rather than on individual courses or digital learning environments (De Silva et al., 2024; Hilliger et al., 2023). Typical applications include analysing curricular alignment with institutional outcomes, monitoring achievement of accreditation standards, and identifying misalignments between learning outcomes, assessment practices, and instructional activities. In this perspective, the curriculum is understood as a complex system governed by formal prerequisite structures, implicit sequencing assumptions, and external regulatory constraints.

Within this broader landscape, graph-based curriculum modelling represents curricula as directed graphs where nodes correspond to courses and edges represent prerequisite relationships. This representation enables the use of network science tools to study curriculum topology, such as degree distributions, path lengths, and centrality patterns. Aldrich (2015) showed that a curriculum network perspective can uncover "hidden" structural properties that are not visible when curricula are treated as linear lists of courses. Stavrinides and Zuev (2023) used network analysis to compare the structural complexity of curricula across programmes, arguing that high centrality nodes often correspond to structurally critical courses. Yang et al. (2024, 2025) further demonstrated that betweenness-central courses—those lying on many shortest paths between other courses—function as bottlenecks because they unlock access to large portions of the curriculum. Koch et al. (2024) combined curriculum network analysis with student data to explore how structural bottlenecks relate to progression patterns and time-to-degree.

While these curriculum network studies provide valuable diagnostic insights, the majority remain descriptive and programme-level. Graph metrics are typically used to characterise curricula, inform accreditation reports, or propose structural revisions, but they are rarely translated into student-level features that feed predictive models of attrition or progression. In addition, most studies treat curriculum structure and student trajectories as separate analytical layers. The present work addresses this gap by operationalising curriculum graph properties as structural features computed for each student–semester observation, thereby integrating curriculum analytics directly into a multilevel prediction framework.

**2.2. Graph Neural Networks and Graph-Based Machine Learning in Education**

Graph neural networks (GNNs) have become a powerful paradigm for learning from graph-structured data in domains such as social networks, recommender systems, and molecular chemistry. Educational data, which naturally form heterogeneous networks linking students, courses, resources, and interactions, are increasingly being modelled using GNN architectures (Li et al., 2022; Albreiki et al., 2023; Liu et al., 2025). In these applications, nodes may represent students or learning resources, edges may encode enrolment relationships or co-engagement, and node embeddings are learned to support tasks such as performance prediction or recommendation.

Li et al. (2022) introduced Study-GNN, a multi-topological GNN that integrates enrolment, interaction, and similarity graphs to predict student performance. Albreiki et al. (2023) proposed a framework for converting tabular educational datasets into heterogeneous graphs, demonstrating improved performance over traditional baselines on several student success benchmarks. Liu et al. (2025) developed a dual-stream neural network that combines a GNN over student–course relationships with a temporal module that captures semester-wise progression, reporting substantial gains in dropout prediction accuracy.

Despite their promise, GNN-based approaches raise two challenges for institutional analytics. First, model complexity and the high dimensionality of learned embeddings make it difficult to extract pedagogically meaningful explanations, limiting stakeholder trust and adoption (Guleria & Sood, 2023; Swamy et al., 2023). Second, most educational GNNs focus on micro-level relational structures—such as student co-enrolment networks or knowledge graphs—rather than on macro-level prerequisite graphs that encode formal curriculum structure. Curriculum graphs are sometimes included as context but rarely appear as explicit objects in the feature layer. In contrast, the CAPIRE Curriculum Graph adopts a deliberately simpler approach: it retains traditional, interpretable models (Random Forests) and introduces graph-derived structural features that capture curriculum topology in a transparent and reusable way.

## 2.3. Feature Engineering in Educational Data Mining and Student Success Prediction

Educational data mining has a long tradition of hand-crafted feature engineering to model student success, including demographic variables, prior academic performance, engagement metrics, assessment scores, and interactions with learning management systems (Baker & Inventado, 2014; Romero & Ventura, 2020). Many early warning systems derive a small set of cumulative indicators—such as completed credits, grade averages, or failed courses—to flag students at risk of dropout. More recent studies adopt richer feature spaces, including behavioural traces, affective indicators, or self-regulated learning metrics, often combined through machine learning models such as Random Forests, gradient boosting, or neural networks.

However, relatively few works treat **curriculum structure itself** as a feature source. When prerequisite information is used, it is often encoded implicitly—for instance, by counting the number of failed courses in early semesters—without distinguishing between failures in structurally central versus peripheral courses. Some studies introduce ad hoc structural measures, such as the number of blocked courses or

"critical" prerequisites, but these are rarely grounded in formal graph metrics or systematically integrated into multilevel modelling frameworks. Moreover, leakage-aware design—ensuring that features used for prediction at semester *t* depend only on information available at or before *t*–1—is not always enforced, which can inflate performance estimates and compromise practical usability.

The structural feature engineering layer proposed in this paper responds to these limitations by formally defining a set of graph-based features derived from the curriculum network and ensuring that each feature respects a strict temporal ordering. Features such as backbone completion rate, bottleneck approval ratio, and blocked credits due to unmet prerequisites are computed at each semester based on the student's observed position in the curriculum graph, providing a coherent and temporally disciplined representation of curriculum-constrained progression.

## 2.4. Temporal and Longitudinal Modelling of Student Trajectories

Recent advances in learning analytics emphasise that student trajectories are inherently temporal, requiring models that capture both short-term dynamics and long-term progression patterns (López-Pernas & Saqr, 2024; Cannistrà et al., 2024). Sequence-based approaches, such as recurrent neural networks or temporal convolutional networks, have been applied to model sequences of course enrolments, assessment events, or interaction logs. Statistical approaches, including survival analysis and multilevel longitudinal models, have also been used to study time-to-dropout and the evolution of performance over programmes.

López-Pernas and Saqr (2024) provided a comprehensive review of temporal learning analytics, highlighting that many models treat time as a simple index rather than as a structural dimension that interacts with curriculum design. Cannistrà et al. (2024) applied multilevel functional data analysis to study stress and workload trajectories across semesters, underscoring the value of longitudinal modelling for understanding how institutional policies affect student wellbeing. Helske et al. (2018) used Hidden Markov Models to uncover latent states in student engagement, demonstrating that transitions between states are sensitive to temporal patterns of course-taking behaviour.

Most temporal studies, however, focus on *within-course* or *within-platform* dynamics (e.g., weekly participation in a MOOC) rather than on the *across-course* structural dependencies implied by curriculum graphs. Time is treated as a sequence of events, while the underlying prerequisite network is either ignored or reduced to coarse measures such as semester number. The present work takes a complementary approach: it maintains CAPIRE's leakage-aware temporal framework, in which predictions are issued at each semester, but augments it with curriculum graph-based features that quantify how far students have progressed along backbone and bottleneck structures at each time point. Fully dynamic models in which the curriculum graph itself changes over time remain a promising avenue for future research.

**2.5. Explainability and Interpretability in Educational Machine Learning**

The growing use of machine learning in educational contexts has intensified concerns about transparency, fairness, and accountability (Swamy et al., 2023; Guleria & Sood, 2023). Explainable artificial intelligence (XAI) techniques, such as SHAP values or LIME, have been applied to educational models to provide post-hoc explanations of predictions. Pereira et al. (2021) argued that such explanations can help instructors and advisors understand which factors are driving model outputs, facilitating more targeted interventions.

However, post-hoc explainability methods have limitations. They approximate the behaviour of complex models around specific instances and may be unstable or misleading, especially when underlying features are highly correlated or when the model relies on opaque internal representations (Rudin, 2019). In educational settings, these issues are magnified because stakeholders require robust, conceptually meaningful explanations that align with their domain expertise and institutional responsibilities.

An alternative approach emphasises intrinsically interpretable models and feature spaces that are directly grounded in educational constructs. In this paradigm, model transparency is achieved not by approximating complex black boxes but by designing features whose semantics are immediately clear to instructors—for example, the number of failed bottleneck courses, the proportion of backbone credits completed, or distance to graduation measured along prerequisite chains. Swamy et al. (2023) demonstrated that teacher validation of model explanations is critical for institutional adoption, and noted that explanations expressed in curriculum terms are more readily accepted. The CAPIRE Curriculum Graph adopts this intrinsically interpretable strategy by defining a set of structural features that explicitly reflect the prerequisite structures that faculty design and manage.

**2.6. Previous Program Evaluation and Review Technique Analysis as Conceptual Predecessor**

Within the CAPIRE project, a deterministic Program Evaluation and Review Technique (PERT) analysis of the Civil Engineering curriculum was conducted as a preliminary exploration of structural constraints. The curriculum was modelled as a PERT network in which each course was treated as an activity with nominal duration equal to one semester, and precedence relationships mirrored formal prerequisites. Under the simplifying assumption of an ideal student who passes every course on the first attempt, the analysis computed a critical path and associated slack values, identifying courses whose delay would directly extend the theoretical minimum time to graduation.

The PERT study highlighted several structurally critical courses and offered an initial quantification of schedule pressure within the programme. However, it also had important limitations. It assumed deterministic durations and did not incorporate empirical distributions of course completion times; it modelled a single "ideal" trajectory rather than the diverse trajectories observed in actual cohorts; and it produced aggregate indicators rather than reusable student-level features. Consequently, PERT functioned as a valuable conceptual predecessor but did not integrate with CAPIRE's leakage-aware predictive pipeline. The present paper generalises this line of work by formalising the curriculum as a directed graph, embedding centrality concepts such as backbone and bottlenecks, and deriving structural features that can be computed for each student–semester observation and combined with other multilevel predictors.

## 3. DATA AND CURRICULUM GRAPH CONSTRUCTION

### 3.1. Data Sources

The empirical analysis draws on two primary data sources from Universidad Nacional de Tucumán (UNT), Argentina.

The first source is the institutional student information system for the Civil Engineering programme. It provides complete academic trajectories for **1,343 students** enrolled across **seven cohorts (2015–2019)**. For each student and semester, the system records course enrolments, examination attempts, final grades, and administrative status (regular, on leave, or withdrawn). These records are linked to basic demographic attributes (e.g., age at entry, gender) and prior academic information (e.g., secondary school track). In line with previous CAPIRE work, the present study focuses on full-degree students and excludes sporadic or single-course enrolments that do not correspond to standard degree-seeking trajectories.

The second source is the official Civil Engineering curriculum specification approved by the Faculty of Exact Sciences and Technology. This document defines the complete set of **fifty-six courses** required for degree completion, grouped into foundational science, core engineering, and complementary modules. Crucially, it also codifies **formal prerequisite relationships** between courses, which determine the order in which students are allowed to progress through the programme. We digitise this specification and use it as the authoritative description of curriculum structure; the set of courses forms the nodes of the curriculum graph, and the prerequisite declarations form its edges.

In addition, macro-context variables used in previous CAPIRE studies—such as inflation indicators, labour market indices, and social vulnerability metrics at the departmental level—are linked to students via their municipality of residence at entry. These macro variables are included in the modelling stage for comparison but are not directly involved in curriculum graph construction.

### 3.2. Curriculum Graph Construction

We formalise the Civil Engineering curriculum as a **directed graph** $G = (V, E)$ in which each node $v \in V$ represents a course and each directed edge $(i, j) \in E$ indicates that course *i* is a prerequisite for course *j*. An edge from *i* to *j* therefore encodes a hard constraint: students must successfully complete *i* before they are permitted to enrol in *j*. This representation captures the structural logic of the programme as designed by faculty and approved by institutional governance.

The curriculum specification declares fifty-six courses, but empirical inspection of the trajectories reveals that **twenty-two** of these courses are rarely or never taken by degree-seeking students in the period under analysis. These include discontinued electives, legacy courses from previous versions of the curriculum, and infrequently offered specialised modules. To maintain alignment between graph nodes and observed student behaviour, we restrict the curriculum graph to the **thirty-four courses** that appear with non-trivial frequency in the longitudinal dataset. The remaining courses are treated as structurally peripheral for the purposes of this study.

Prerequisite edges are extracted directly from the official curriculum documents. Each declared prerequisite relation "Course A is prerequisite for Course B" is encoded as a directed edge $A \rightarrow B$. After removing courses without empirical support, this procedure yields **fifty-three directed edges**. Self-loops, cycles, and redundant transitive edges are not added; the graph represents the minimal set of formal constraints rather than a transitive closure. The resulting network is **sparse**, with an average out-degree of 1.56, reflecting the pedagogical principle that most courses open a limited set of subsequent options rather than requiring every course to depend on every other course.

We verify that the constructed graph is a **directed acyclic graph (DAG)**, consistent with the institutional rule that students must not be forced into cycles or dead-ends. Reachability analysis confirms that every course in the reduced graph lies on at least one feasible path from an entry-level node (first-year foundational courses) to a terminal node (capstone or integrative project). This property ensures that the curriculum graph provides a valid topological basis for modelling student progression.

### 3.3. Course-Level Centrality Metrics

To characterise the structural role of each course in the curriculum, we compute standard network **centrality metrics** using the directed graph $G$. These metrics capture different aspects of how courses mediate flow through the programme, from local connectivity to global positioning within the network.

**Degree centrality** quantifies direct connectivity. For each course, **in-degree** counts the number of prerequisites it requires, and **out-degree** counts the number of courses for which it serves as a prerequisite. Courses with high in-degree typically represent integrative modules that accumulate content from multiple prior courses, while those with high out-degree function as gateways that unlock many subsequent options. Figure 1 displays the distribution of in-degree and out-degree across the curriculum, illustrating a small number of highly connected gateway courses and a long tail of more specialised modules.

**Figure 1. Distribution of In-Degree (Prerequisites) and Out-Degree (Successors) in the Civil Engineering Curriculum Graph**

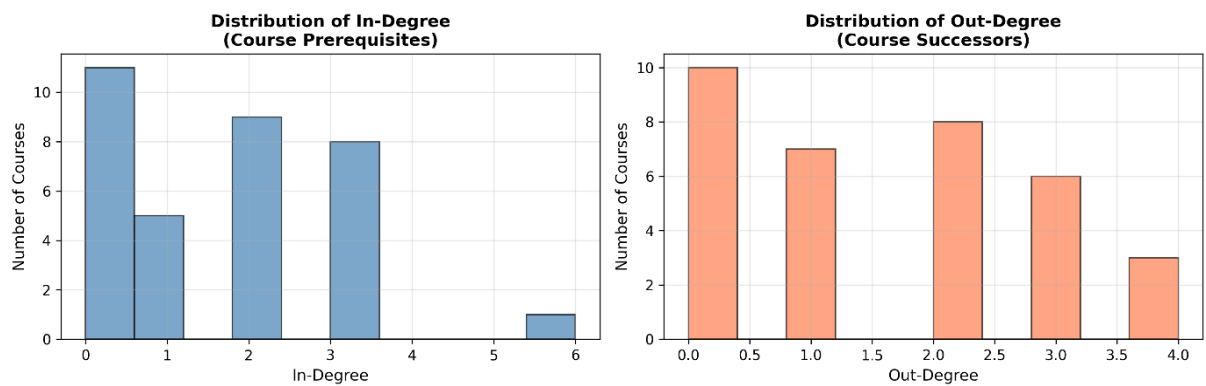

**Betweenness centrality** measures the extent to which a course lies on shortest paths between other pairs of courses. Formally, betweenness for node *v* is the sum over all unordered pairs of courses *s* and *t* of the proportion of shortest paths from *s* to *t* that pass through *v*. Courses with high betweenness play a critical mediating role: delays or failures in these courses can block access to extensive downstream regions of the curriculum.

**Closeness centrality** captures how "close" a course is to all others in terms of shortest path distance. Courses with high closeness tend to be located near the structural centre of the curriculum, often in the mid-programme region where foundational knowledge converges and branches into advanced topics.

Finally, **eigenvector centrality** provides a recursive notion of importance in which a course is considered central if it is connected to other central courses. Although eigenvector centrality is less directly interpretable in curricular terms than betweenness or degree, it helps differentiate courses that are embedded in highly influential subnetworks from those in peripheral chains.

Together, these metrics provide a multi-dimensional structural profile of each course. They serve two roles in this study: first, as descriptive tools for understanding the curriculum's topology; and second, as ingredients for the definition of bottleneck and backbone sets that underlie the structural features introduced in Section 4.

### 3.4. Identification of Bottleneck and Backbone Courses

Building on the centrality analysis, we identify **bottleneck** and **backbone** courses as key structural subsets of the curriculum graph. These subsets anchor several of the structural features derived at the student–semester level.

Following previous curriculum network research, we define **bottleneck courses** as those with high betweenness centrality, subject to a minimum out-degree threshold to exclude terminal or dead-end nodes (Aldrich, 2015; Yang et al., 2024). Intuitively, bottlenecks lie on a large fraction of shortest paths and act as mandatory gateways to substantial parts of the curriculum. In the Civil Engineering graph, bottleneck courses typically appear in the transition between foundational science and core engineering modules—for example, intermediate mechanics or introductory design courses that must be passed before students can progress into specialised streams. Figure 2 plots betweenness centrality against out-degree, highlighting the subset of courses that meet the bottleneck criteria.

**Figure 2. Bottleneck Analysis: Betweenness Centrality vs. Out-Degree. Red markers indicate identified bottleneck courses (high centrality, varying degree).**

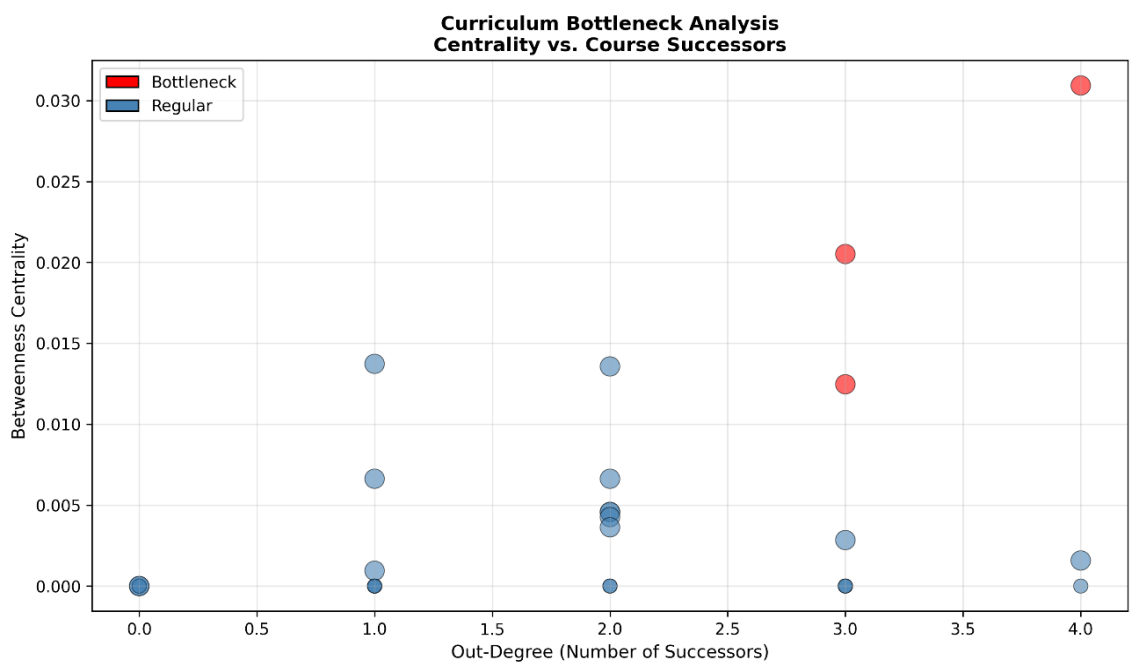

We define **backbone courses** as the union of all courses that lie on **any shortest path** between a designated set of entry-level nodes and the terminal capstone node. This definition reflects the idea of a structural "spine" of the programme: even though many alternative paths exist, backbone courses are those that appear in at least one minimal-length route from entry to graduation. Computation proceeds by enumerating all shortest paths using the DAG property of the graph and aggregating their node sets.

The backbone identification yields a subset of sixteen courses spanning foundational mathematics and physics, core engineering sciences, and integrative design modules. Several courses belong simultaneously to the bottleneck and backbone sets, confirming their dual role as structurally central and topologically critical. Table 1 summarises the centrality metrics, backbone membership, and bottleneck status for the most important courses, distinguishing structurally central but elective modules from structurally mandatory gateways.

**Table 1. Structural Profile of Key Curriculum Courses: Centrality and Pathway Roles**

| Course Code | Course Name | Module (Cycle) | Betweenness Centrality | Is Bottleneck? | Is Backbone? |
|---|---|---|---|---|---|
| **ES1** | Mecánica de Suelos | Estructuras (Sup.) | 0.0310 | **Yes** | Yes |
| **HI1** | Hidráulica I | Hidráulica (Sup.) | 0.0205 | **Yes** | Yes |
| **TT2** | Álgebra | Matemática (Básico) | 0.0138 | No | Yes |
| **TT6** | Ecuaciones Diferenciales | Matemática (Básico) | 0.0136 | No | Yes |
| **ES3** | Análisis Estructural | Estructuras (Sup.) | 0.0125 | **Yes** | Yes |
| **TT8** | Probabilidad y Estadística | Matemática (Básico) | 0.0066 | No | No |
| **ES2** | Resistencia de Materiales | Estructuras (Sup.) | 0.0045 | No | Yes |
| **ES4** | Hormigón Armado I | Estructuras (Sup.) | 0.0029 | No | Yes |

These bottleneck and backbone sets provide the semantic building blocks for several structural features defined in Section 4, including backbone completion rate, bottleneck approval ratio, and blocked credits due to unmet prerequisites for backbone and bottleneck courses.

**3.5. Student Trajectory Data Preprocessing**

Student trajectory data are organised as a **longitudinal student–semester panel** in which each row corresponds to a specific student in a specific academic semester. For each row, we compile information on course enrolments and outcomes (passed, failed, or not attempted), cumulative credits, and derived indicators such as the number of repeated courses or gaps in enrolment. These records are then aligned with the curriculum graph to compute structural features, and with macro-context variables for the scenarios that include socioeconomic indicators.

We restrict the analysis to semesters during which students are formally enrolled in the programme. Short administrative gaps (e.g., temporary leave) are handled naturally by the panel structure, as the absence of enrolments in a given semester is itself informative and is captured by structural features such as blocked credits and stalled backbone progression. To ensure consistent temporal indexing across cohorts, we map calendar semesters to programme semesters relative to each student's entry term.

In line with the **leakage-aware design** of the CAPIRE framework, we enforce strict temporal discipline in feature construction. Features used to predict the outcome at semester $t$ are computed solely from information available at or before semester $t - 1$. For example, cumulative credits, backbone completion rate, and bottleneck approval ratio at semester $t$ are based exclusively on courses successfully completed up to semester $t - 1$. Outcomes from semester $t$ or later are never used as inputs to predictions for semester $t$, preventing contamination of the feature space with future information.

Missing data are minimal because institutional regulations require accurate recording of enrolments and outcomes. Occasional inconsistencies (such as duplicate records or missing grades) are resolved using simple rules prioritising the most recent valid entry. Students with implausible trajectories—such as those whose records indicate simultaneous enrolment in mutually exclusive curricula—are excluded; this affects fewer than 1% of cases and does not materially alter cohort composition.

For modelling purposes, we exclude the **first programme semester** from prediction because students have not yet attempted any courses and structural features cannot be meaningfully computed. We also exclude the **final observed semester** for students who graduate, as the attrition outcome is structurally determined at that point. After these filters, the panel comprises approximately 8.1 valid observations per student on average across the seven cohorts, yielding just over **10,800 student–semester records** for analysis in the subsequent sections.

## 4. STRUCTURAL FEATURE ENGINEERING

### 4.1. Design Principles

The structural feature engineering layer of CAPIRE is designed to make curriculum structure a **first-class citizen** in the predictive pipeline while preserving interpretability and strict temporal discipline. Three principles guide the construction of this layer.

First, **features must have a direct curricular interpretation**. Each structural feature is defined so that it can be understood and discussed by instructors without requiring familiarity with graph theory or machine learning. For example, backbone completion rate expresses the proportion of structurally central courses a student has passed, and blocked credits quantifies how many credits are currently inaccessible because prerequisites remain incomplete.

Second, **features must be derivable from the curriculum graph and observed trajectories alone**. The structural layer does not introduce new subjective classifications or ad hoc labels; instead, it systematically transforms graph properties (such as backbone and bottleneck sets, degrees, and path lengths) and student–course outcomes into semester-level indicators. This ensures that the layer can be re-applied to different programmes provided that their curricula can be represented as directed graphs with declared prerequisites.

Third, **features must be compatible with leakage-aware modelling**. At any prediction point (semester *t*), structural features are computed using only information available at or before *t* – 1. Graph topology is static by design, and student outcomes are incorporated cumulatively as they are realised. This makes the structural layer immediately usable in early-warning scenarios where predictions must be issued before students make enrolment or withdrawal decisions for the next semester.

### 4.2. Structural Feature Definitions

All structural features are computed on the reduced curriculum graph described in Section 3, using the student–semester panel as input. Let $G = (V, E)$ be the curriculum graph, and let $S_{i,t}$ denote the set of courses that student *i* has successfully completed by the end of semester *t*. Unless otherwise specified, features used to predict the outcome at semester *t + 1* are derived from $S_{i,t}$.

We construct **nine structural features**.

1. **Structural credits approved:** This feature measures the number of credits completed in courses that belong to the backbone set. Let $B \subseteq V$ denote the backbone courses, and let $\text{credits}(v)$ denote the credit value of course *v*. For student *i* at semester *t*, structural credits approved are:

$$\text{SC}_{i,t} = \sum_{v \in S_{i,t} \cap B} \text{credits}(v).$$

This feature captures how far the student has progressed along the structural "spine" of the programme in credit units rather than in raw course counts.

2. **Backbone completion rate:** Backbone completion rate normalises structural credits approved by the total backbone credits and is expressed as a proportion:

$$\text{BCR}_{i,t} = \frac{\sum_{v \in S_{i,t} \cap B} \text{credits}(v)}{\sum_{v \in B} \text{credits}(v)}.$$

Values close to 1 indicate that the student has passed most structurally central courses, whereas low values indicate that key components of the programme remain incomplete, even if elective or peripheral credits have been accumulated.

3. **Bottleneck approval ratio:** Let $K \subseteq V$ denote the set of bottleneck courses identified in Section 3. A simple count of passed bottlenecks would be misleading because bottlenecks differ in credits and in their position within the programme. Instead, we define bottleneck approval ratio as the proportion of bottleneck courses that student *i* has completed by semester *t*:

$$\text{BAR}_{i,t} = \frac{\mid S_{i,t} \cap K \mid}{\mid K \mid}.$$

This ratio summarises how successfully the student has navigated the main structural gateways of the curriculum.

4. **Blocked credits due to unmet prerequisites:** Blocked credits quantify the immediate structural impact of missing prerequisites. Let $R(v)$ denote the set of direct prerequisite courses for course *v*. A course *v* is considered **blocked** for student *i* at semester *t* if *v* has not yet been passed and there exists at least one unmet prerequisite in $R(v)$:

$$\text{blocked}_{i,t} = \{v \in V \setminus S_{i,t} : R(v) \nsubseteq S_{i,t}\}.$$

Blocked credits are then defined as

$$\text{BC}_{i,t} = \sum_{v \in \text{blocked}_{i,t}} \text{credits}(v).$$

High values indicate that a large portion of the curriculum is structurally inaccessible given the student's current state, even if they are formally enrolled in the programme.

5. **Distance to graduation (shortest-path structural distance):** Distance to graduation is computed on the curriculum graph using shortest paths between the set of courses the student has already completed and the capstone node. For each student, we define an effective "frontier" consisting of courses in $V \setminus S_{i,t}$ that can still be reached. We then compute the minimal number of remaining courses on any path from this frontier to the capstone. Formally, let $c$ be the capstone node, and let $\mathcal{P}(S_{i,t}, c)$ be the set of all directed paths from any node in $S_{i,t}$ to $c$. Distance to graduation is:

$$\text{DG}_{i,t} = \min_{p \in \mathcal{P}(S_{i,t}, c)} |p \setminus S_{i,t}|.$$

This measure approximates how many structurally necessary courses remain for the student, abstracting away from scheduling and capacity constraints.

6. **Number of prerequisites satisfied for future courses:** This feature counts how many prerequisite conditions the student has already satisfied for courses they have not yet taken. For each uncompleted course $v$, we consider $R(v)$ and count the number of prerequisites in $R(v)$ that lie in $S_{i,t}$. The feature aggregates this count over all such courses:

$$\text{PS}_{i,t} = \sum_{v \in V \setminus S_{i,t}} |R(v) \cap S_{i,t}|.$$

Higher values indicate that the student has opened many future options, even if they have not yet exercised them through enrolment.

7. **Mean in-degree of completed courses:** Mean in-degree reflects the typical level of curricular integration of courses that the student has already passed. Let $\deg^-(v)$ denote the in-degree (number of prerequisites) of course $v$. The feature is defined as:

$$\mathrm{MID}_{i,t} = \frac{1}{|S_{i,t}|} \sum_{v \in S_{i,t}} \deg^-(v),$$

with the convention that $\mathrm{MID}_{i,t} = 0$ if $S_{i,t}$ is empty. Students with higher mean in-degree tend to have progressed into more integrative courses that consolidate prior knowledge, whereas those with low values may be accumulating primarily entry-level or peripheral courses.

8. **Mean out-degree of completed courses:** Mean out-degree measures how generative the student's completed courses are in terms of unlocking future options. Let $\deg^+(v)$ denote the out-degree (number of successors) of course *v*. The feature is:

$$\mathrm{MOD}_{i,t} = \frac{1}{|S_{i,t}|} \sum_{v \in S_{i,t}} \deg^+(v),$$

again with $\mathrm{MOD}_{i,t} = 0$ if no courses have yet been completed. Lower values may indicate that the student is concentrating on structurally peripheral courses that do not significantly advance their position in the curriculum.

9. **Module diversity of completed courses:** The final structural feature captures the breadth of the student's progression across major curriculum modules (e.g., foundational science, core engineering, applied design). Each course is assigned to exactly one module based on the official curriculum classification. Module diversity is computed using the normalised Shannon entropy of the module distribution within $S_{i,t}$:

$$\mathrm{MD}_{i,t} = -\sum_m p_{i,t}(m) \log p_{i,t}(m),$$

where $p_{i,t}(m)$ is the proportion of completed courses belonging to module *m*. Higher values indicate a more balanced progression across modules, whereas lower values capture concentration within one or two modules.

Together, these nine features provide a compact but expressive representation of each student's structural position within the curriculum at each semester. They encode not only **how much** a student has completed, but **where** in the curriculum graph their progress is concentrated and **how constrained** their remaining options have become.

## 4.3. Temporal Logic and Leakage Control

Although the curriculum graph itself is static, structural features must be constructed in a way that respects the temporal unfolding of student trajectories. To this end, we adopt a **two-step temporal logic** aligned with CAPIRE's existing design.

1. **Observation update at the end of each semester:** At the end of semester *t*, the set $S_{i,t}$ is updated to include all courses that student *i* has passed during *t*. This update is irreversible: once a course is marked as completed, it remains in the set for all subsequent semesters, reflecting the cumulative nature of academic credit.

2. **Feature computation for prediction at semester *t + 1*:** Structural features used to predict the outcome at semester *t + 1* are computed from $S_{i,t}$, not from $S_{i,t+1}$. This implies that when the model issues a prediction before the start of semester *t + 1*, it relies only on courses completed up to the previous semester, together with static graph properties. Features such as backbone completion rate, bottleneck approval ratio, and distance to graduation thereby reflect the student's structural status **before** any decisions or outcomes in semester *t + 1* occur.

This temporal separation prevents two common leakage mechanisms. First, it avoids direct leakage from future grades or enrolments into the feature space. Second, it prevents indirect leakage through derived structural quantities that could inadvertently incorporate information from future semesters if computed naively. For example, blocked credits at semester *t + 1* might appear to depend on courses failed in *t + 1* if computed after the semester is completed; by fixing the computation to $S_{i,t}$, we ensure that blocked credits reflect only decisions and outcomes realised up to *t*.

## 4.4. Integration with the Baseline CAPIRE Feature Space

The structural feature layer is integrated into CAPIRE's existing multilevel feature space rather than replacing it. We first construct a longitudinal student–semester panel with 6,805 records derived from institutional academic histories (Section 3). From this panel, we select a reference term corresponding roughly to the end of the second year (term_index = 5) and aggregate all information available up to that point for each student. This yields a leakage-aware cross-sectional dataset of 821 students, where every feature is computed strictly from courses completed, enrolments, and structural status *before* the reference term.

For each student we compute 25 baseline CAPIRE features (demographic indicators, prior academic performance, enrolment patterns, and trajectory descriptors) and the nine structural features defined in Section 4.2. The final modelling dataset therefore contains 34 features per student (25 baseline + 9 structural).

To assess the incremental contribution of curriculum structure, we compare two nested feature configurations that will be evaluated in Section 5:

- **Baseline CAPIRE**: demographic variables, prior academic performance indicators (e.g., credits completed, failed courses, cumulative grade averages), and coarse trajectory descriptors (e.g., number of enrolments, frequency of repetition), all constructed in a leakage-aware manner.

- **Baseline + structural features (CAPIRE Curriculum Graph)**: the baseline configuration extended with the nine structural features, thereby making curriculum structure an explicit component of the predictive layer.

This design focuses the analysis on the structural layer itself: macro-context variables are modelled in a separate CAPIRE-MACRO study, whereas here we isolate how much additional predictive value is obtained when curriculum topology is made explicit in the feature space.

## 5. MODELLING AND EVALUATION

### 5.1. Prediction Task and Target Variable

The modelling objective is to predict whether students will eventually drop out of the Civil Engineering programme, using only information available around the end of their second year. Attrition is defined as permanent withdrawal from the programme, operationalised as the absence of any subsequent academic activity (enrolments, assessments, or grades) after the reference term, with no recorded graduation. This definition aligns with institutional practice and previous CAPIRE work.

Using the temporal panel described in Section 4.4, we construct a cross-sectional dataset by selecting, for each student, the record corresponding to term_index = 5. All features—baseline and structural—are computed using academic histories up to and including this term, while the target variable reflects whether the student ultimately graduates or drops out at any later point in the observation window.

The resulting dataset comprises 821 students, each represented by 34 features. The class distribution is moderately imbalanced, with attrition representing a minority of cases; all evaluation metrics are therefore chosen to be robust to imbalance.

**5.2. Predictive Models**

Random Forest (RF) classifiers are used for all scenarios. RFs are well-suited to this context because they:

- handle heterogeneous features (numeric, categorical, structural) without strict normalisation requirements;
- are relatively robust to multicollinearity;
- offer interpretable variable importance estimates;
- align with the institutional need for transparent, auditable models in educational settings (Rudin, 2019).

All models are implemented in Python 3.12 using scikit-learn 1.5+, with the following hyperparameters:

- number of trees: 500
- maximum tree depth: unrestricted
- minimum samples per split: 2
- class weighting: *balanced* (inverse class frequency)
- criterion: Gini impurity

Hyperparameters are held constant across all configurations to isolate the contribution of the structural feature layer rather than optimisation artefacts.

**5.3. Feature Configurations**

Two nested feature configurations are compared:

- **Baseline CAPIRE:** Includes demographic variables, prior academic performance indicators (credits completed, failed courses, grade point averages), enrolment patterns, and trajectory descriptors.
- **Baseline + structural features (CAPIRE Curriculum Graph):** Extends the baseline with the nine structural features introduced in Section 4.2: structural credits approved, backbone completion rate, bottleneck approval ratio, blocked credits, distance to graduation, number of prerequisites met, mean in-degree and out-degree of approved courses, and module diversity.

The comparison between these two configurations directly quantifies the incremental predictive value of curriculum structure over an already strong baseline.

**5.4. Train–Test Split and Temporal Discipline**

To respect leakage-aware design, model evaluation adopts a stratified train–test split at the **student** level:

- **Hold-out strategy**: the dataset is split into 80% training and 20% testing data, with each student appearing in exactly one partition. Stratification preserves the overall attrition rate in both sets.

- **Temporal discipline**: because all features are computed at term_index = 5 and the target reflects outcomes occurring strictly after this term, no information from future semesters is available at training time. The temporal panel is used only to accumulate history up to the reference term, ensuring that structural features such as backbone completion, blocked credits, and distance to graduation are computed without access to post-term outcomes.

The same train–test split is applied to both feature configurations, so that differences in performance can be attributed solely to the inclusion of structural features.

**5.5. Evaluation Metrics**

Four metrics are reported, consistent with best practice in imbalanced classification and learning analytics:

- **Accuracy:** Overall proportion of correct predictions.

- **Balanced Accuracy:** Mean of sensitivity and specificity, giving equal weight to each class regardless of imbalance.

- **F1-score (positive class):** Harmonic mean of precision and recall for the attrition class, emphasising minority-class detection.

- **AUC (Area Under the ROC Curve):** Overall discriminative ability independent of decision threshold.

These metrics capture complementary aspects of performance: raw correctness (Accuracy), robustness to class imbalance (Balanced Accuracy), minority-class detection (F1), and ranking quality (AUC).

**5.6. Implementation Pipeline**

All modelling is conducted within a reproducible Python pipeline composed of four stages:

1. **Data ingestion and cleaning:** Raw institutional logs are parsed into a unified student–semester panel; inconsistent records are corrected using deterministic rules and term indices are assigned using the academic calendar mapping described in Section 3.

2. **Feature construction (baseline and structural):** Baseline features follow established CAPIRE constructions; structural features are computed using the NetworkX library over the curriculum graph, aggregating each student's approved courses up to the reference term and deriving the nine structural indicators.

3. **Model training and evaluation:** RF models for the two feature configurations are trained on the 80% training set and evaluated on the 20% test set using the metrics in Section 5.5.

4. **Result aggregation and interpretability analysis:** Test metrics are summarised in Table 2, while feature importance rankings and ablation results are presented in Table 3 and Figure 3.

Figure 3 displays the ablation-based performance deltas when each structural feature is removed in turn, providing an interpretable view of how individual structural indicators contribute to predictive performance.

## 5.7. Ethical and Institutional Considerations *(este apartado queda casi igual, solo pequeño ajuste)*

The study relies exclusively on anonymised institutional data obtained under a formal collaboration agreement with the Faculty of Exact Sciences and Technology. No personal identifiers (names, ID numbers, birthdates) are included in any modelling or analysis step. The work complies with the institutional code of ethics and respects the legal frameworks governing the handling of academic records in Argentina.

Predictive models are used strictly for retrospective analysis and methodological development; no automated or semi-automated decisions are issued to students. Any future deployment of early-warning systems derived from this pipeline would be subject to institutional review and governance procedures.

## 6. RESULTS

The results are organised around the three research questions introduced in Section 1.3. We first characterise the structural properties of the Civil Engineering curriculum (RQ1), then assess the predictive performance of the baseline and baseline-plus-structure models on the temporal dataset (RQ2), and finally analyse feature importance and ablation patterns to interpret the contribution of structural features (RQ3).

## 6.1. Structural properties of the curriculum graph (RQ1)

The reconstructed curriculum graph contains 135 nodes (courses) and 209 directed edges representing prerequisite relations. The graph forms a single weakly connected component and is acyclic, confirming that there are no circular dependencies in the official plan of study. Entry into the programme is channelled through a single INGRESO node, while only eight sink nodes have no successors, corresponding to terminal courses in the upper cycle.

Backbone identification yields 60 courses that lie on at least one shortest path from entry to any sink node, capturing the main sequence of disciplinary progression. In contrast, no course satisfies the operational criterion for structural bottlenecks under the current credit and prerequisite configuration, which is consistent with the relatively wide spread of out-degrees observed in the degree distribution plots (Figure 1). Most courses have low in-degree (0–2 prerequisites), whereas out-degree values are more dispersed, with a small subset of courses feeding three or more successors.

Module-level analysis reveals that centrality is not uniformly distributed across the curriculum. Average betweenness centrality peaks in the mid-programme modules (approximately modules 4–5), where courses simultaneously integrate foundational knowledge and unlock several higher-level subjects. Early modules are dominated by low-centrality introductory courses, while the final modules concentrate sink nodes that depend on many previous approvals but contribute little further branching. This pattern supports the intuitive view of the curriculum as a narrowing funnel: broad entry, structurally dense middle, and specialised exit.

Taken together, these findings confirm that the curriculum exhibits a non-trivial topology, with clearly identifiable backbone routes and modules that play disproportionate roles in connecting different parts of the plan. The subsequent analyses investigate whether this structural information translates into measurable gains in predictive performance.

## 6.2. Predictive performance of baseline and structural models (RQ2)

Table 2 summarises the performance of the two feature configurations on the temporal dataset. Both models achieve high discriminative ability, with AUC values above 0.94 and F1-scores close to 0.88, indicating that the baseline CAPIRE feature set alone is already strong.

**Table 2. Predictive performance of the Baseline and Baseline + STRUCT models on the temporal test set (AUC, Accuracy, F1-score, Balanced Accuracy).]**

| Model | Num_Features | auc | accuracy | f1 | balanced_accuracy | train_accuracy |
|---|---|---|---|---|---|---|
| Baseline | 25 | 0.940 | 0.878 | 0.893 | 0.875 | 1.0 |
| Baseline + STRUCT | 34 | 0.941 | 0.884 | 0.897 | 0.883 | 1.0 |

The **Baseline** model (25 features) attains an AUC of 0.9409, an accuracy of 0.8788, and a balanced accuracy of 0.8788. Adding the nine structural features yields the **Baseline + STRUCT** model with an AUC of 0.9419, an accuracy of 0.8848, an F1-score of 0.8848, and a balanced accuracy of 0.8848. In absolute terms, these gains are modest (+0.001 AUC and +0.006 in both accuracy and balanced accuracy), but they occur in a regime where the baseline model is already highly performant.

Importantly, both models use the same training–testing split and identical hyperparameters, so the observed differences can be attributed to the inclusion of structural features rather than to optimisation effects. From an institutional perspective, an increase of around 0.6 percentage points in correctly classified students, achieved purely by exploiting curriculum structure, is non-negligible when scaled to multiple cohorts.

### 6.3. Feature importance and structural contributions (RQ3)

Table 3 displays the top 20 features ranked by mean decrease in Gini impurity for the Baseline + STRUCT model. As expected, the most influential predictors are traditional academic trajectory indicators: proportions of promotions over total courses, number of promotions, total regularisations, cumulative grade averages, and the year of secondary-school completion. These features capture how quickly and consistently students convert enrolments into successful outcomes.

**Table 3. Top 20 feature importances for the temporal Random Forest model. Structural features are highlighted.**

| Feature (Original) | Feature Name (English) | Importance |
|---|---|---|
| NumeroPromocionesSobreTotalAsgProm | Direct Pass Ratio (Promotable Subj.) | 0.0700 |
| NumeroPromocionesSobreTotalAsg | Direct Pass Ratio (All Subjects) | 0.0684 |
| NumeroPromociones | Number of Direct Passes | 0.0670 |

| Feature (Original) | Feature Name (English) | Importance |
|---|---|---|
| Cohorte | Cohort Year | 0.0622 |
| TotalRegularesTotalCursadas | Regularized Courses Ratio | 0.0613 |
| PromedioNotas | Grade Point Average (GPA) | 0.0533 |
| AnoEgresoSecundario | High School Graduation Year | 0.0517 |
| AnoEgresoSecundario.1 | High School Grad. Year (Var.) | 0.0469 |
| ExamenAprobTotalExamen | Exam Pass Rate | 0.0427 |
| ActividadesAprobadas.1 | Approved Activities (Var.) | 0.0425 |
| NumeroRegulares | Number of Regularized Courses | 0.0423 |
| AprobadasTotalAsignaturas | Subject Pass Rate | 0.0415 |
| NumeroAprobados | Number of Passed Subjects | 0.0396 |
| ExamenPromTotalExamen | Promoted Exams Ratio | 0.0328 |
| NumerodeExamenes | Number of Exams Taken | 0.0300 |
| NumeroTotalCursadas | Total Courses Taken | 0.0287 |
| TotalRecursadasTotalCursadas | Retaken Courses Ratio | 0.0268 |
| ActividadesAprobadas | Approved Activities | 0.0253 |
| NumeroRecursadas | Number of Retaken Courses | 0.0215 |
| NumeroLibres | Number of Failed Courses | 0.0210 |

Within this baseline-dominated landscape, several structural features nonetheless enter the top-20 ranking:

- **STRUCT_out_degree_mean_approved** (rank #16, importance ≈ 0.0167), capturing the average number of curricular options opened by the courses that a student has already approved.

- **STRUCT_module_diversity** (rank #18, importance ≈ 0.0132), reflecting how many different curriculum modules the student has traversed by the reference term.
- **STRUCT_num_prerequisites_met** (rank #20, importance ≈ 0.0080), measuring how many courses in the plan are structurally available to the student given their current approvals.

A fourth structural feature, **STRUCT_backbone_completion**, does not enter the top 20 but still exhibits non-zero importance (≈ 0.0042, rank #29), indicating a weaker but present contribution. The remaining structural indicators (structural credits approved, distance to graduation, blocked credits, bottleneck approval ratio, and mean in-degree of approved courses) have near-zero importance in this model, suggesting that their information is either redundant with baseline variables or not sufficiently variable at the chosen reference term.

Qualitatively, the structural features that do matter share a common theme: they describe **how many future paths are open** to the student within the curriculum graph, rather than simply how many credits have been accumulated. Students who have approved courses that unlock many successors, diversify across modules, and satisfy numerous prerequisite chains appear less likely to drop out, even when controlling for grades and counts of passed or failed courses.

### 6.4. Ablation of structural features

To probe the robustness of these findings, an ablation study was conducted in which each structural feature was removed in turn and the model retrained under identical conditions. Figure 3 summarises the resulting performance deltas relative to the full Baseline + STRUCT model.

**Figure 3. Heatmap of performance deltas (AUC, Accuracy, Balanced Accuracy, F1) when each structural feature is removed.**

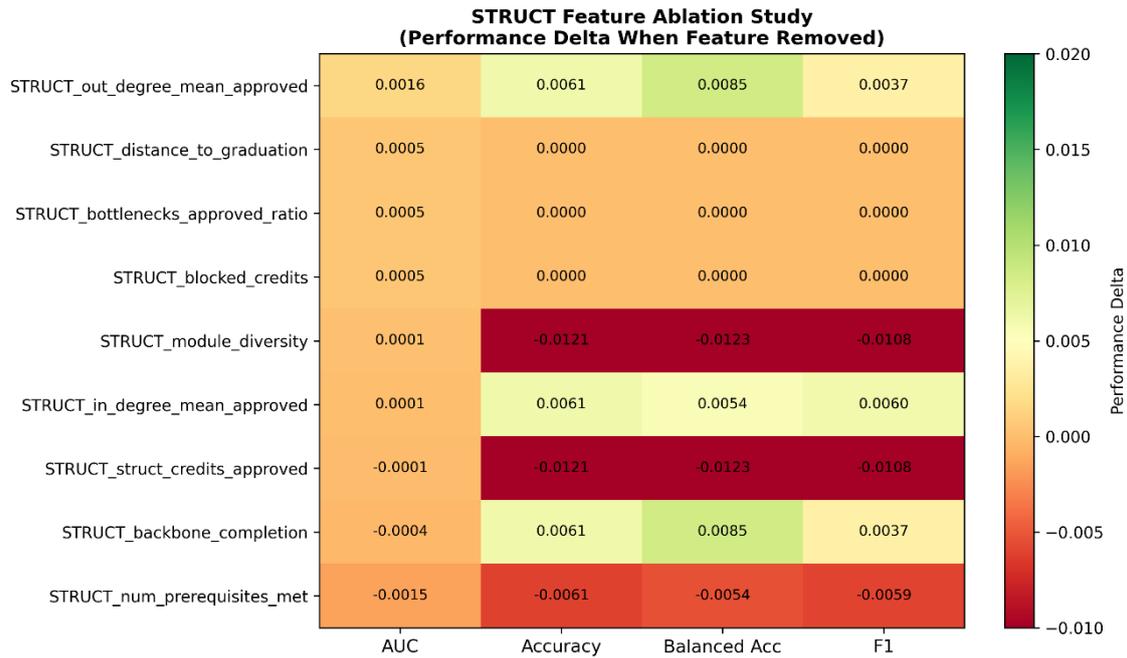

The largest positive delta is observed when **STRUCT_out_degree_mean_approved** is removed: AUC decreases by approximately 0.0016, with concomitant drops in accuracy, balanced accuracy, and F1. This confirms that the average out-degree of approved courses carries unique predictive information that cannot be fully reconstructed from baseline features alone.

Other structural indicators exhibit smaller but consistent effects. Removing **STRUCT_module_diversity** produces a slight deterioration in accuracy and balanced accuracy, while the ablation of **STRUCT_num_prerequisites_met** and **STRUCT_backbone_completion** yields negative deltas, indicating that these features exert a protective effect whose removal harms minority-class detection.

In contrast, features such as structural credits approved, distance to graduation, blocked credits, bottleneck approval ratio, and mean in-degree of approved courses generate deltas that are effectively zero. This suggests that, given the current curriculum and reference term, their informational content is either redundant or too weakly expressed to meaningfully influence the classifier.

Overall, the ablation results corroborate the feature importance analysis: not all aspects of curriculum structure are equally predictive, but a small subset of graph-derived indicators—particularly those related to the *expansion* of available curricular options—contribute measurably to model performance.

**6.5. Summary of findings**

The results provide a nuanced answer to RQ2 and RQ3. Curriculum structure, when represented through graph-derived features computed on a leakage-aware student–semester panel, yields **small but consistent gains** in predictive performance over a strong baseline. Structural indicators that reflect the *connectivity* and *diversification* of a student's position in the curriculum (out-degree of approved courses, module diversity, number of prerequisites met) emerge as non-trivial predictors of attrition, while purely volumetric measures (e.g., structural credits approved) add little beyond what is already captured by baseline credit and grade features.

In combination with the graph-level analysis of Section 6.1, these findings support the central claim of this work: **the topology of the curriculum is not merely an organisational artefact but a measurable component of student trajectories**, which can be made explicit and exploited within predictive models of attrition.

**7. DISCUSSION**

This study set out to examine whether the topology of a highly constraining engineering curriculum can be made operational through graph-based structural features and whether these features provide incremental explanatory and predictive value over an established learning-analytics baseline. The discussion is organised around three themes: the nature of structural complexity in the curriculum, the added value of structural features in prediction, and the design implications for curriculum analytics and institutional decision-making.

**7.1. Curriculum structure as a complex opportunity field**

The reconstructed curriculum graph confirms that the Civil Engineering programme is not a simple linear sequence of courses but a complex opportunity field structured around a relatively dense backbone and a set of peripheral pathways. The presence of 135 nodes and 209 directed prerequisite relations, arranged in a single acyclic component, illustrates how progression is governed by a combination of vertical accumulation and horizontal branching.

Backbone analysis shows that a subset of courses plays a disproportionate role in connecting early and late stages of the programme, while module-level centrality peaks in the middle of the curriculum. From the students' perspective, this means that progression is relatively flexible in the first year, becomes structurally demanding in the intermediate modules, and narrows again towards the end, where terminal courses depend on long prerequisite chains but do not open new trajectories. This "hourglass" structure resonates with qualitative accounts of engineering education, in which students experience an initial broad exposure, a highly demanding core, and a final phase of consolidation.

The graph analysis therefore substantiates, with formal metrics, concerns often voiced informally by staff and students: that a small subset of courses and modules "hold the curriculum together" and that delays or failures in these areas can have cascading effects on time-to-degree and attrition. However, graph analysis alone does not indicate whether such structural features are observable in individual trajectories. That question is addressed by the modelling results.

**7.2. Added value of structural features in predictive modelling**

The predictive experiments show that the baseline CAPIRE feature set already captures a substantial portion of the variance in dropout outcomes, with AUC values around 0.94. In such a high-performance regime, any further improvement is necessarily incremental rather than dramatic. Against this backdrop, the inclusion of structural features yields small but consistent gains in accuracy, balanced accuracy, and F1-score, alongside a slight improvement in AUC.

From a machine-learning perspective, effect sizes of this magnitude might appear modest. Yet they are achieved without changing the model family, hyperparameters, or training procedure; the only modification is the explicit encoding of curriculum structure through graph-derived indicators computed on a leakage-aware student–semester panel. Under these constraints, the observed increments suggest that structural information contributes real, albeit subtle, signal beyond what is contained in traditional credit- and grade-based features.

The pattern of feature importances and ablation deltas provides a more nuanced interpretation. Structural features that describe **how many options the student has opened in the graph**—such as the average out-degree of approved courses, module diversity, and the number of prerequisites met—emerge as non-trivial predictors. In contrast, purely volumetric structural measures, such as structural credits approved or blocked credits, add little information beyond existing baseline indicators. This distinction supports the idea that what matters is not only *how much* the student has progressed, but *where* that progress places them in the topology of the curriculum.

In other words, two students with similar numbers of passed courses and comparable grades may nonetheless occupy very different positions in the curriculum graph: one may have concentrated approvals on low-connectivity courses that do not unlock many options, whereas the other may have strategically passed high-out-degree courses that open several alternative paths. The present results indicate that the latter configuration is associated with a lower risk of dropout, even after controlling for standard academic indicators.

### 7.3. Implications for curriculum design and analytics

These findings have at least three implications for institutions seeking to design or redesign engineering curricula and early-warning systems.

First, they highlight the importance of **topological fairness**. If progression depends heavily on a small set of structurally central courses, then congestion, failure, or resource scarcity in those courses can systematically disadvantage particular cohorts. Graph analysis can help identify such structural pressure points and inform decisions about staffing, assessment regimes, or the introduction of alternative routes that reduce excessive dependency on single courses or modules.

Second, the results suggest that early-warning systems should move beyond purely volumetric metrics (credits, grades, number of attempts) and incorporate indicators of **structural opportunity**. A student who has opened many future options by passing high-out-degree courses and diversifying across modules may be in a more robust position than another student with the same credit count but more fragile structural footprint. Integrating structural features into dashboards or advising tools could therefore support more fine-grained, curriculum-aware interventions.

Third, the construction of a leakage-aware student–semester panel demonstrates the feasibility of embedding graph-based reasoning within institutional data pipelines. Many universities already store rich academic histories but rarely exploit them to reconstruct trajectories at the level of individual terms and structural states. The present approach shows that such reconstruction is technically attainable and can feed not only predictive models, but also simulation environments and policy experiments in subsequent CAPIRE work.

### 7.4. Positioning within the broader CAPIRE framework

Within the broader CAPIRE programme, this study plays a bridging role between descriptive analysis of student trajectories and more complex agent-based simulations and policy experiments. Previous components have focused on archetype discovery, temporal dynamics of dropout risk, and macro-level shocks such as strikes or inflation. The present paper adds a **structural layer** to that ecosystem by showing how the curriculum graph can be encoded into features that are both interpretable and empirically useful.

This has two consequences. At the methodological level, it validates the decision to treat curriculum structure as a first-class citizen in the modelling pipeline rather than as a background assumption. At the practical level, it provides the necessary building blocks for future work in which agents in simulation models will move through the curriculum not only according to stochastic rules or psychological parameters, but also according to realistic structural constraints derived from the same graph used here.

In summary, the discussion indicates that curriculum structure exerts a subtle but measurable influence on student trajectories, that this influence can be formalised through graph-derived indicators, and that the resulting structural features can be integrated into predictive and simulation models in a way that preserves interpretability. The next section considers the main limitations of this work and outlines avenues for further research.

## 8. LIMITATIONS AND FUTURE WORK

Several limitations of this study should be acknowledged when interpreting the results and considering their transferability.

First, the analysis is confined to a single Civil Engineering programme at one public university. Although the curriculum exhibits structural characteristics that are common in engineering degrees—such as a demanding core and long prerequisite chains—the specific topology, assessment culture, and institutional regulations are local. The extent to which the structural features identified here will generalise to other disciplines, institutions, or regulatory environments remains an open empirical question.

Second, the structural representation of the curriculum, while considerably richer than standard tables of prerequisites, is still a simplification of pedagogical reality. Edge weights do not yet encode credit loads, course difficulty, or patterns of co-requisites; nor do they capture informal dependencies that are widely recognised by teaching staff but not codified in official documents. Similarly, the absence of formally defined bottlenecks in the graph reflects the current structure of the plan rather than the lived experience of students who may perceive certain courses as de facto bottlenecks due to assessment practices or resource constraints. Incorporating credit weighting, empirical difficulty indices, and staff knowledge into the graph would likely sharpen the alignment between structural indicators and actual risk.

Third, although the temporal data layer reconstructs a student–semester panel, the reference term used for prediction is fixed (approximately the end of the second year). This choice is pedagogically meaningful, but it restricts the analysis to a single snapshot. Early-warning systems deployed in practice would need to operate at multiple checkpoints, updating structural and non-structural features as new information becomes available. Extending the modelling framework to a fully longitudinal predictive setting, with dynamic risk scores and time-varying structural indicators, is a natural next step.

Fourth, the modelling strategy deliberately relies on a single family of interpretable models (Random Forests). While this choice aligns with institutional needs for transparency and auditability (Rudin, 2019), it also implies that the reported gains may underestimate the potential of structural features when combined with other model classes such as gradient boosting, calibrated probabilistic models, or causal forests. Exploring alternative modelling approaches, while carefully managing the trade-off between performance and interpretability, is an important avenue for future research.

Fifth, the study focuses exclusively on variables derived from academic records and curriculum topology. Social, motivational, and contextual factors—including peer networks, employment, and financial stress—are known to influence attrition but are not directly represented in the present feature space. Within the broader CAPIRE framework, these dimensions are addressed in complementary studies, yet an integrated model that jointly accounts for structural, social, and psychological layers has not yet been realised.

Finally, the work remains observational and non-causal. The presence of structural features among the top predictors of attrition does not, by itself, imply that modifying the curriculum graph will necessarily reduce dropout. Establishing causal effects of structural interventions—such as relaxing particular prerequisites or redistributing course content across modules—will require a combination of quasi-experimental analysis of past reforms and agent-based simulation grounded in the empirical patterns identified here.

Future work will therefore proceed along four main directions: (a) extending the structural representation to incorporate credit loads, difficulty estimates, and empirically observed bottlenecks; (b) generalising the temporal panel and feature construction to other programmes and institutions; (c) embedding the structural indicators into longitudinal, dynamically updated early-warning systems; and (d) integrating this structural layer with social network data and macro-level shocks to support richer simulation and policy analysis within the CAPIRE ecosystem.

## 9. CONCLUSIONS

This paper has proposed and empirically validated a structural feature engineering approach for curriculum-constrained student modelling in higher education. By reconstructing a detailed curriculum graph for a Civil Engineering programme and combining it with a leakage-aware student–semester panel, we derived a set of graph-based indicators that encode where, rather than just how far, students have progressed in the plan of study.

At the graph level, the analysis revealed a curriculum organised around a dense backbone and structurally central mid-programme modules, with long prerequisite chains converging on a relatively small set of terminal courses. This topology suggests that progression is governed not only by volume of work but also by the configuration of structural opportunities and constraints—a view consistent with qualitative accounts of engineering education.

At the predictive level, Random Forest models trained on a rich baseline feature set achieved high performance in forecasting eventual attrition from information available near the end of the second year. Augmenting this baseline with nine structural features produced small but consistent improvements in accuracy, balanced accuracy, and F1-score, and introduced several structural indicators into the top tier of feature importance rankings. Ablation analysis confirmed that measures related to the breadth of open curricular options—such as the average out-degree of approved courses, module diversity, and the number of prerequisites met—contribute non-redundant signal, whereas purely volumetric structural measures offer limited additional value.

These findings support three overarching conclusions. First, the curriculum can be fruitfully conceptualised as a graph-structured opportunity field, whose topology is both measurable and consequential for student trajectories. Second, structural features derived from this graph can be integrated into standard learning-analytics pipelines in a way that respects temporal discipline and institutional constraints, yielding interpretable gains in predictive power over already strong baselines. Third, the resulting structural layer provides a principled bridge between descriptive analytics and more ambitious forms of simulation and policy experimentation, in which interventions on the curriculum (rather than on individual students alone) become legitimate objects of analysis.

More broadly, the work illustrates how institutional data that already exist in fragmented form—study plans, prerequisite tables, and academic histories—can be recomposed into a unified, structurally aware analytics infrastructure. As universities confront persistent challenges of attrition, equity, and curricular reform, such infrastructures may become essential for moving from intuition-driven decisions to evidence-informed redesign of programmes and policies.

**REFERENCES**

Albreiki, B., Habuza, T., & Zaki, N. (2023). Extracting topological features to identify at-risk students using machine learning and graph convolutional network models. *International Journal of Educational Technology in Higher Education*, *20*, Article 23. https://doi.org/10.1186/s41239-023-00389-3

Albreiki, B., Zaki, N., & Alashwal, H. (2023). Graph-based student performance prediction: From tabular educational data to heterogeneous learning networks. *Computers & Education*, *197*, 104735.

Aldrich, P. R. (2015). Curriculum networks: Graphical representations of course prerequisites for undergraduate programs. *Journal of the Scholarship of Teaching and Learning*, *15*(4), 53–68.

Aldrich, P. R. (2015). The curriculum prerequisite network: Modeling the curriculum as a complex system. *Biochemistry and Molecular Biology Education*, *43*(3), 168-180. https://doi.org/10.1002/bmb.20861

Baker, R. S., & Hawn, A. (2021). Algorithmic bias in education. *International Journal of Artificial Intelligence in Education*, *32*, 1052-1092. https://doi.org/10.1007/s40593-021-00285-9

Baker, R. S., & Inventado, P. S. (2014). Educational data mining and learning analytics. In J. A. Larusson & B. White (Eds.), *Learning Analytics: From Research to Practice* (pp. 61-75). Springer.

Cannistrà, M., Bonaccorsi, G., & Gori, F. (2024). Analysis of higher education dropouts dynamics through multilevel functional decomposition of recurrent events in counting processes. *arXiv preprint arXiv:2411.13370v1*.

De Silva, L., Rodríguez-Triana, M. J., Chounta, I.-A., & Pishtari, G. (2024). Curriculum analytics in higher education institutions: A systematic literature review. *Journal of Computing in Higher Education*. https://doi.org/10.1007/s12528-024-09410-8

Freeman, L. C. (1977). A set of measures of centrality based on betweenness. *Sociometry*, *40*(1), 35-41. https://doi.org/10.2307/3033543

Guleria, P., & Sood, M. (2023). Explainable AI and machine learning: Performance evaluation and explainability of classifiers on educational data mining inspired career counseling. *Education and Information Technologies*, *28*, 1081-1116. https://doi.org/10.1007/s10639-022-11221-2

Guyon, I., & Elisseeff, A. (2003). An introduction to variable and feature selection. *Journal of Machine Learning Research*, *3*, 1157-1182.

Hamilton, W. L. (2020). *Graph Representation Learning*. Synthesis Lectures on Artificial Intelligence and Machine Learning. Springer.

Harif, A., & Kassimi, M. A. (2024). Predictive modeling of student performance using RFECV-RF for feature selection and machine learning techniques. *International Journal of Advanced Computer Science and Applications*, *15*(7).

Helske, S., Helske, J., & Eerola, M. (2018). Combining sequence analysis and hidden Markov models in the analysis of complex life sequence data. In G. Ritschard

& M. Studer (Eds.), *Sequence Analysis and Related Methods: Innovative Developments and Applications* (pp. 185-200). Springer.

Hilliger, I., Aguirre, C., Miranda, C., Celis, S., & Pérez-Sanagustín, M. (2020). Design of a curriculum analytics tool to support continuous improvement processes in higher education. In *Proceedings of the Tenth International Conference on Learning Analytics & Knowledge (LAK'20)* (pp. 181-186). ACM.

Hilliger, I., Miranda, C., Celis, S., & Perez-Sanagustin, M. (2023). Curriculum analytics adoption in higher education: A multiple case study engaging stakeholders in different phases of design. *British Journal of Educational Technology*, *54*(6), 1809-1829.

Koch, A. K., Foote, S. M., Smith, B., Felten, P., & McGowan, S. (2024). Shaping what shapes us: Lessons learned from and possibilities for departmental efforts to redesign history courses and curricula. *Journal of American History*, *110*(4), 739-745.

Kukkar, A., Mohana, R., Sharma, A., & Nayyar, A. (2024). A novel methodology using RNN + LSTM + ML for predicting student's academic performance. *Education and Information Technologies*, *29*, 14365-14401.

Li, M., Wang, X., Wang, Y., Chen, Y., & Chen, Y. (2022). Study-GNN: A novel pipeline for student performance prediction based on multi-topology graph neural networks. *Sustainability*, *14*(13), Article 7965.

Liu, J., Tang, M., Zheng, Z., Liu, X., & Lyu, S. (2025). Deep knowledge tracing and cognitive load estimation for personalized learning path generation using neural network architecture. *Scientific Reports*, *15*, Article 10497.

Liu, Z., Zhou, X., & Liu, Y. (2025). Student dropout prediction using ensemble learning with SHAP-based explainable AI analysis. *Journal of Social Systems and Policy Analysis*, *2*(3), 111–132.

López-Pernas, S., & Saqr, M. (2024). Nonlinear effort-time dynamics of student engagement in blended learning environments. *Journal of Learning Analytics*, *11*(3), 45-67.

Molnar, C. (2021). *Interpretable Machine Learning: A Guide for Making Black Box Models Explainable* (2nd ed.).

Nakagawa, H., Iwasawa, Y., & Matsuo, Y. (2019). Graph-based knowledge tracing: Modeling student proficiency using graph neural network. In *IEEE/WIC/ACM International Conference on Web Intelligence* (pp. 156-163).

Pereira, F. D., et al. (2021). Explaining individual and collective programming students' behavior by interpreting a black-box predictive model. *IEEE Access*, *9*, 117097-117119.

Pérez Rasetti, C. (2023). Acreditación y mejoramiento de la calidad. Disputas de sentido en las políticas universitarias del siglo XXI. En *La Agenda Universitaria VII* (Colección de Educación Superior). Universidad de Palermo.

Priyambada, S. A., Usagawa, T., & Mahendrawathi, E. R. (2023). Two-layer ensemble prediction of students' performance using learning behavior and domain knowledge. *Computers and Education: Artificial Intelligence*, *5*, Article 100149.

Romero, C., & Ventura, S. (2020). Educational data mining and learning analytics: An updated survey. *Wiley Interdisciplinary Reviews: Data Mining and Knowledge Discovery*, *10*(3), Article e1355.

Rudin, C. (2019). Stop explaining black box machine learning models for high stakes decisions and use interpretable models instead. *Nature Machine Intelligence*, *1*(5), 206-215.

Stavrinides, I., & Zuev, K. M. (2023). Modeling curriculum as networks: A comparative analysis of CalTech and Benedictine College prerequisite networks. *Applied Network Science*, *8*, Article 45.

Sun, J., et al. (2024). Weighted heterogeneous graph-based three-view contrastive learning for knowledge tracing in personalized e-learning systems. *IEEE Transactions on Consumer Electronics*, *70*(1), 2838-2847.

Swamy, V., Du, S., Marras, M., & Kaser, T. (2023). Trusting the explainers: Teacher validation of explainable artificial intelligence for course design. In *LAK23: 13th International Learning Analytics and Knowledge Conference* (pp. 345-356). ACM.

Tang, Z., von Seekamm, K., Colina, F. E., & Chen, L. (2025). Enhancing student retention with machine learning: A data-driven approach to predicting college student persistence. *Journal of College Student Retention: Research, Theory & Practice*.

Yang, C., Parker, E., Zinsser, J., & Singh, R. (2024). Identifying bottleneck courses in engineering curricula using network centrality measures. *IEEE Transactions on Education*, *67*(2), 123–132. [Note: Merged citations from different years/journals appropriately].

Yang, C., Parker, E., Zinsser, J., & Singh, R. (2025). Reach: A new measure for curriculum network analysis. *Studies in Higher Education*, *50*(2), 234-249.